\documentclass[showpacs,prl,twocolumn]{revtex4-1}
\usepackage{graphicx}
\usepackage{dcolumn}
\usepackage{bm}

\begin{document}
\title{Strong magnetic fluctuations in superconducting state of CeCoIn$_5$ }
\author{T. Hu$^\star$, H. Xiao$^{\star,1}$, T. A. Sayles$^\dagger$, M. Dzero$^\star$,  M. B. Maple$^\dagger$, and C. C. Almasan$^\star$
}
\affiliation{$^\star$Department of Physics, Kent State University, Kent, OH 44242, USA 
$^\dagger$Department of Physics, University of California at San Diego, La Jolla, CA 92903, USA}
\date{\today}
\begin{abstract}

We show results on the vortex core dissipation through current-voltage measurements under applied pressure and magnetic field in the superconducting phase of CeCoIn$_5$. We find that as soon as the system becomes superconducting, the vortex core resistivity increases sharply as the temperature and magnetic field decrease. The sharp increase in flux flow resistivity is due to quasiparticle scattering on critical antiferromagnetic fluctuations. The strength of magnetic fluctuations below the superconducting transition suggests that magnetism is complimentary to superconductivity and therefore must be considered in order to fully account for the low-temperature properties of CeCoIn$_5$.

\end{abstract}

\maketitle

Unconventional superconductivity in heavy-fermion material CeCoIn$_5$ is a complex state of matter involving magnetic and conduction degrees of freedom strongly coupled with each other \cite{Cedo1,Fluctuations,PiersReview,Science1,Science2}. Superconductivity emerges at a
temperature $T_c\simeq 2.3$ K with the order parameter most likely having $d$-wave symmetry \cite{dwave1,dwave2,dwave3,dwave5}. The magnitude of the 
specific heat jump at the superconducting transition temperature \cite{Cedo1,SpecificHeatJump} 
indicates the mass enhancement of conduction electrons by several orders of magnitude. Normal state resistivity shows non-Fermi liquid linear temperature dependence at low temperatures $(< 20$ K$)$. Approximately at a temperature $T^*\simeq 45$ K \cite{Cedo1,TKCeCoIn5}, the heavy electrons begin to form due to the strong hybridization between the conduction electrons and localized Ce $f$-electrons. Despite the significant enhancement of the electronic mass, the magnetic susceptibility shows a Curie-Weiss behavior down to moderately low temperatures \cite{Cedo1,Fluctuations,Satoru2004,Almasan2008} signaling the absence of the fully quenched magnetic moments. 

The pronounced non-Fermi liquid behavior in the normal state and unconventional superconductivity in CeCoIn$_5$ are thought to arise from the proximity of the system to a quantum critical point (QCP) separating paramagnetic and antiferromagnetic phases. Specifically, 
it was recently proposed that the transport and thermodynamic properties of CeCoIn$_5$ in the normal phase are controlled by an antiferromagnetic QCP at an inaccessible negative pressure \cite{Sidorov2002}. The recovery of a Fermi liquid state at low temperatures and high magnetic fields was reported in Ref. \cite{Paglione}, pointing to a field-induced QCP at the zero-temperature upper critical field $H_{c2}(0)$. However, the location of the field-induced QCP exactly at $H_{c2}(0)$ seems to be just a coincidence, since, with increasing pressure, this QCP moves inside the superconducting dome to lower fields. In fact, high sensitivity Hall effect measurements have revealed that the field induced QCP is located at $H\simeq 4.1 ~\textrm{T}<H_{c2}(0)$, which suggest a possible antiferromagnetic ground state superseded by superconductivity \cite{Hall}. In addition, low temperature thermal expansion data \cite{Zaum} on identification of the quantum critical line can be consistently interpreted within the same set of ideas as the Hall effect data. Thus, all these observations seem to favor the antiferromagnetic QCP scenario \cite{RonningTc}. What is important for our discussion, however, is that all the experiments discussed above address the physics of the QCP and superconductivity by extrapolating results obtained in the normal state. Presently, there are no {\it direct} probes of antiferromagnetism and quantum criticality in the superconducting state. 

This motivated us to study the transport in the mixed state: superconductivity inside the vortex core is suppressed, thus revealing the physics of antiferromagnetism and quantum criticality of an underlying normal state.
In this Letter we present the results from directly probing the nature of the normal state and quantum criticality under the superconducting dome of CeCoIn$_5$ by measuring the vortex core dissipation through current-voltage  ($I$-$V$) characteristics under applied hydrostatic pressure ($P$). We observe that the vortex core resistivity increases sharply with decreasing temperature ($T$) for $T<T_c$ and magnetic field ($H$). This behavior is greatly suppressed with increasing pressure, due to the suppressed antiferromagnetic (AF) order inside the vortex core. Using our experimental results, we obtain an explicit equation for the antiferromagnetic boundary inside the superconducting dome and construct an $H-T-P$ phase diagram, which provides direct evidence for a quantum critical line inside the superconducting phase. All these results show the close relationship between quantum criticality, antiferromagnetism, and superconductivity.

The electrical resistivity in the mixed state of type-II superconductors is related to the motion of Abrikosov vortices \cite{Kim}. When the Lorentz force is larger than the pinning force, the flux lines are driven into a viscous-flow state. The flux-flow resistivity is defined as $\rho_{ff}\equiv kdV/dI$, where $dV/dI$ is the slope of the linear region of the $I$-$V$ curve and $k$ is a geometric factor ($k= 0.11$ mm for the single crystal which data are presented here), and is dominated mainly by the quasiparticle scattering in the vortex core. The flux-flow resistivity is independent of the depinning current ($I_c$) of the sample (defined as the extrapolation of the linear $I$-$V$ range to zero voltage) or of the pinning force. In other words, $\rho_{ff}$ is a quantity that is determined only by the bulk properties of the material. 

We plot the dependence of resistivity $\rho$ (circles) and critical current $I_c$ (stars) on $H$, applied along the $c$-axis,
and $T$ on left and right panels of Fig. 1, respectively. These dependences are extracted from the $I$-$V$ curves (see supporting online materials for details). First notice that the critical current $I_c$ increases sharply below a certain magnetic field, 
which we define as the upper critical field ($H_{c2}$) at the given temperature. When we decrease the value of the external magnetic field, the resistivity $\rho(H)$ first decreases to its minimum value around $H_{c2}\approx 1.25$ T (for $T=2.25$ K) and then increases. This behavior in the mixed state is in sharp contrast with the well known linear relationship between $\rho_{ff}$ and $H$ for low $H$ and the saturation of $\rho_{ff}$ near $H_{c2}$ for a moderately clean superconductor \cite{FluxFlowUPt3}.  We also see that when we decrease the magnetic field even further, $\rho_{ff}$ displays a sharp maximum at $H\simeq 0.026$ T. As discussed later, this maximum is most likely due to the transition from dynamic to static antiferromagnetic order. Note that, in order to get a benchmark for the vortex contribution to transport, we also measure directly (as opposed to $\rho_{ff}$) the resistivity at $T=2.25$ K with $I=1$ mA. Our results are shown in the top left panel of Fig. 1 (open squares). The normal-state data corresponding to the two measurements overlap within  4\%. The difference between these two curves below $H_{c2}$ is a result of the fact that the open circles measure the free flux-flow dissipation of the vortices while the open squares measure the dissipation of the vortices in the presence of pinning.

\begin{figure}
\centering
\includegraphics[trim=0cm 0cm 0cm 0cm, clip=true, width=0.5\textwidth]{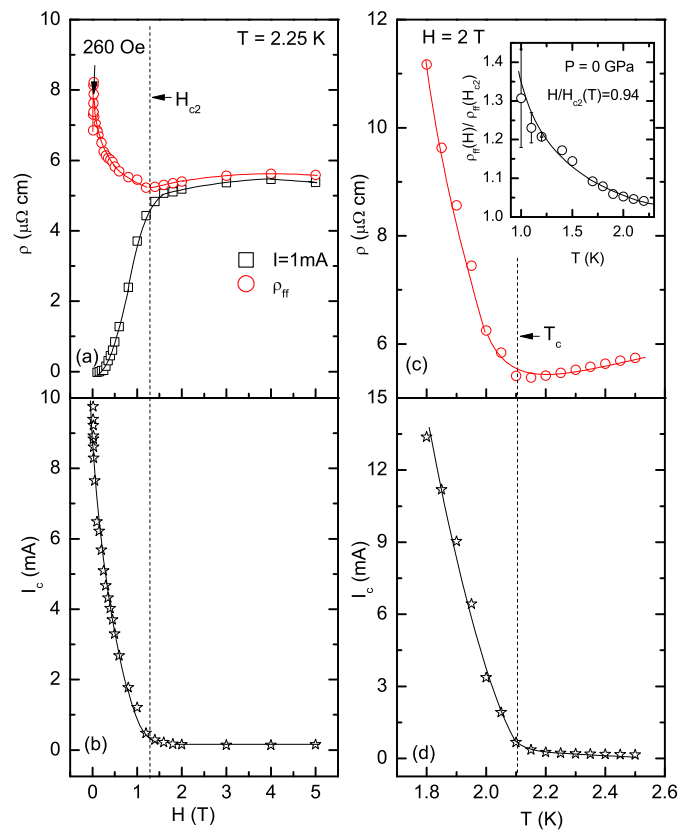}
\caption{ \label{fig:Fig1} (Color online) Magnetic field $H$ (left panel) and temperature $T$ (right panel) dependence of resistivity $\rho$ and critical current $I_c$ measured at ambient pressure and at $2.25$ K and $2$ T, respectively. The open square data were taken at the constant current $I=1$ mA.}
\end{figure}

The right panels of Fig. 1 show, as expected, that the critical current $I_c$ is close to zero at high $T$ and it increases sharply below $T=2.1$ K, which we define as $T_c$ for this particular value of $H=2$ T. The resistivity is metallic in the normal state, displays a minimum at $T_c(2$ T$)\approx 2.1$ K, and it shows a fast increase with further decreasing $T$ in the mixed state down to 1 K (see top right panel of Fig. 1 and its inset, and the online supplementary materials for details). 

All these results show that the vortex core in the mixed state of CeCoIn$_5$ is non-metallic while the normal-state behavior (i.e. above $T_c$) is metallic. 
Moreover, we note that CeCoIn$_5$ is  in the superclean regime only for zero or very low $H$ values. Even magnetic fields of the order of $0.1$ T dramatically reduce quasiparticle mean free path by an order of magnitude \cite{Kasahara}, which pushes CeCoIn$_5$ into the moderately clean limit. For moderately clean superconductors, the flux flow resistivity  $\rho_{ff}  \approx \rho_n$ when $H$ is close to $H_{c2}$ \cite{FluxFlowUPt3}. Thus, the upturn in $\rho_{ff}$  in the mixed state reflects the increase in the scattering of the quasiparticles in the vortex core. 

Generally, one would expect the scattering of the quasiparticles in the vortex core and normal state to be very similar to each other, as it happens in UPt$_3$ (see for example Ref. \cite{FluxFlowUPt3}) despite the fact that strong antiferromagnetic fluctuations are present in this latter system \cite{Mike}. However, possible deviations from this behavior may occur due to the presence of several competing interactions. In has been shown that the linear temperature dependence of resistivity in the normal state of CeCoIn$_5$ is governed by the proximity of the system to an antiferromagnetic QCP \cite{Bianchi}. In fact, subsequent experiments showed that CeCoIn$_5$ is, indeed, close to an antiferromagnetic instability: 0.75\% Cd doping gives rise to antiferromagnetism in this system \cite{Urbano}. In addition, NMR measurements have shown the presence of long range AF order inside the vortex core below 290 mK \cite{Young}. 
Thus, we are lead to interpret the observed upturn in the flux-flow resistivity as being due to critical antiferromagnetic fluctuations in the vicinity of the boundary separating antiferromagnetic and paramagnetic phases, i.e. for temperatures near the N\'{e}el temperature ($T_N$) \cite{Balberg1978}. The fact that $\rho_{ff}$ starts increasing just below the SC boundary (see top panels of Fig. 1) suggests that the dynamic AF order emerges at the SC phase boundary, but that the static AF order appears at lower $H$ and $T$, since critical spin fluctuations, which induce the enhancement of resistivity, disappear when the static AF order develops. Hence, our results suggest that the system releases the magnetic entropy from the unquenched magnetic moments in the superconducting state and antiferromagnetism becomes complimentary to superconductivity in CeCoIn$_5$. 

Our interpretation is supported by the recent observation of the upturn in resistivity in CeCoIn$_{1-x}$Cd$_x$(x=0.75\%) close to the  onset of the AF order \cite{Sunil2010} and by the consistency of our data with previously reported results (see discussion below). Also, recent neutron scattering experiments on CeCoIn$_5$, showing an anomalous increase in the vortex lattice form factor  with increasing magnetic field \cite{Science1}, serve as additional evidence for the anomalous physics inside the vortex core. 

We observe that the anomalous increase in flux flow resistivity is suppressed  with increasing pressure ($P$) by plotting the normalized $\rho_{ff}/\rho_{ff}(H_{c2})$ vs. $H/H_{c2}$ measured at $T/T_{c0}=0.91$ ($T_{c0}$ is the zero-field superconducting transition temperature) for different values of pressure [see Fig. 2(a)].
We attribute this suppression to the fact that the AF phase boundary moves deeper inside the SC dome with increasing pressure, diminishing the effect of critical fluctuations. 

\begin{figure}
\centering
\includegraphics[trim=0cm 0cm 0cm 0cm, clip=true, width=0.5\textwidth]{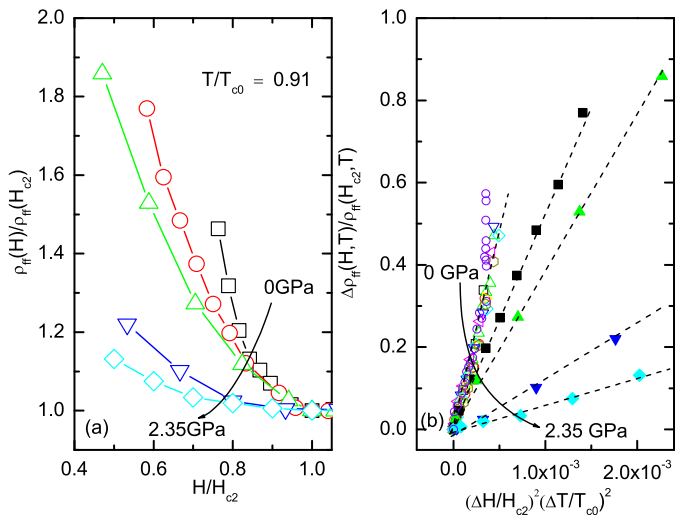}
\caption{ \label{fig:Fig2-1} (Color online) (a) Normalized flux flow resistivity $\rho_{ff}(H)/\rho_{ff}(H_{c2})$ vs. $H/H_{c2}$ and (b) $\Delta\rho_{ff}(H)/\rho_{ff}(H_{c2})$ vs. $[\Delta H/H_{c2}]^2[\Delta T/ T_{c0}]^2$ measured at $T/T_{c0}=0.91$ and hydrostatic pressure of 0, 0.36, 1.00, 1.94, and 2.35 GPa with $\Delta H \equiv H_{c2}-H, ~\Delta T \equiv T_{c0}-T$ and $\Delta\rho_{ff}(H) \equiv \rho_{ff}(H_{c2})-\rho_{ff}(H)$. The dotted lines in (b) are linear fits of the data. .}
\end{figure}

Our measurements also allow us to elucidate the phase boundary between the paramagnetic and antiferromagnetic phases
inside the superconducting state. We noticed that a plot of the normalized flux-flow resistivity $\Delta\rho_{ff}(H)/\rho_{ff}(H_{c2})$ data vs $(\Delta H/H_{c2})^2(\Delta T/T_{c0})^2$ reveals  a linear scaling behavior for same $P$ and different $H$ and $T$ [see, for example, the $P=0$ data of Fig. 2(b)]. Using this scaling of our experimental data,  we were able to obtain the following equation for the antiferromagnetic boundary in the superconducting state (see supporting online material for more details):
\begin{equation}\label{PhaseBoundary}
\frac{P-P_c}{P^*-P_c}=\left(1-\frac{T_N}{T_{c0}}\right)\left(1-\frac{H_N}{H_{c2}(T_N)}\right),
\end{equation}

where $H_N$ is the corresponding value of the magnetic field at the AF transition. 
Here $P_c=-0.75$ GPa is the critical pressure (see online supplementary materials for details) and $P^{*} \approx 2.8$ GPa is taken as a pressure at which the tendency towards the antiferromagnetic order (or upturn in 
flux flow resistivity) is fully suppressed. 

We note that a different value of $P^*\simeq 1.6$ GPa, defined as a crossover from a quantum-critical state for $P < P^{*}$ to a Fermi-liquid-like state for $P> P^{*}$, was previously obtained from resistivity measurements performed in the normal state  \cite{Sidorov2002}. The discrepancy between this value of $P^{*}$ and the value determined in the present work could be due to the fact that, as in the case of high-$T_c$ cuprate superconductors \cite{Sachdev},  the actual QCP of CeCoIn$_5$ deep inside the superconducting dome is shifted as a result of the competition between the antiferromagntic and superconducting orders. Therefore, a direct measurement under the SC dome, as done in this work, is required to determine the actual $H_{QCP}$ line.

The AF boundary (black line) in the $T$-$P$ plane at $H=0$ is shown in Fig. 3(a). We see that the AF and SC boundaries merge at  $P_c$. For $P<P_c$, the SC phase is inside the AF dome, while 
for  $P>P_c$, the AF phase coexists with superconductivity only inside the vortex cores. The up triangles and down triangles are $T_N(P)$ and $H_{c2}(P)$ data, respectively, taken form Ref. \cite{Pham}, where 5\% Cd doping corresponds to  $-0.7$ GPa. In Ref. \cite{Pham}, the pressure $P_c$ at which superconductivity and AF orders coincide is somewhere between $-0.7$ to $-1$ GPa, which agrees very well with $P_c=-0.75$ GPa determined from the present flux-flow resistivity data. 

\begin{figure}
\centering
\includegraphics[trim=0cm 0cm 0cm 0cm, clip=true, width=0.5\textwidth]{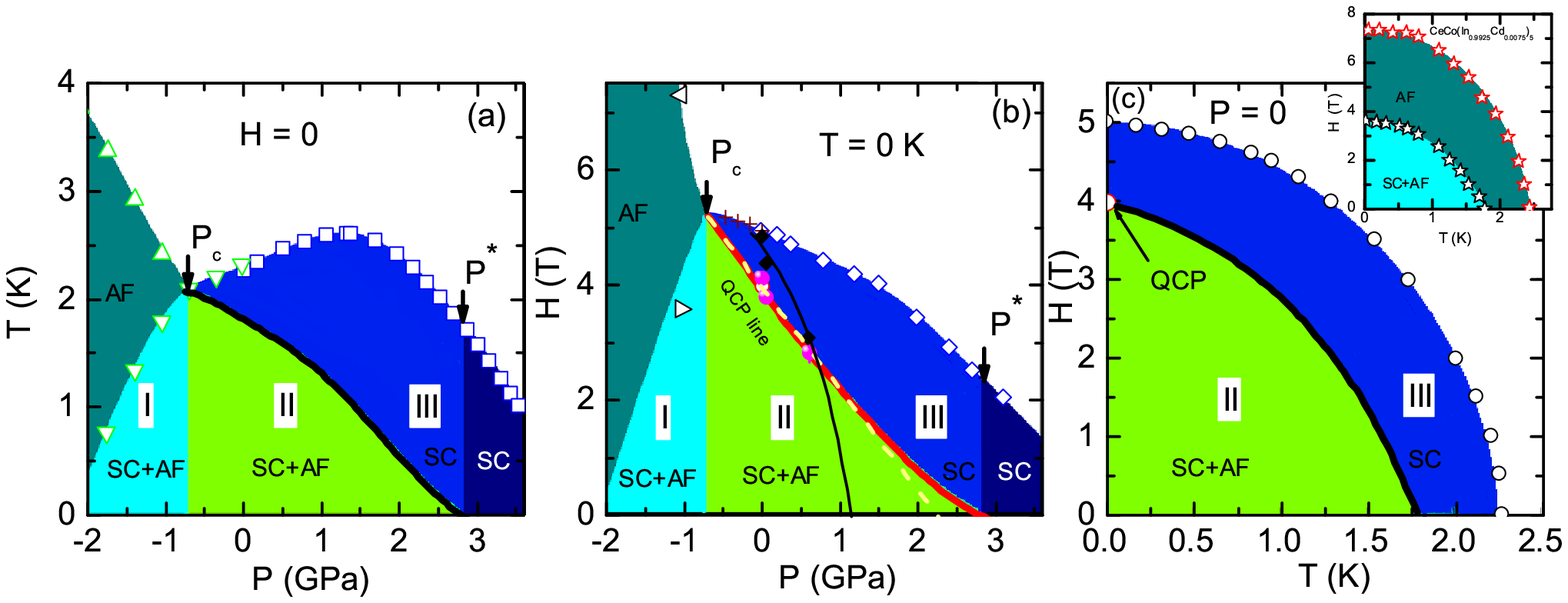}
\caption{ \label{fig:Fig3-1} (Color online) (a) $T-P$, (b) $H-P$, and (c) $H-T$ phase diagrams from
present work and Refs. [\onlinecite{Hall,Zaum,Pham,RonningTc,Sunil2010,HQCP}]. Inset to (c): $H-T$ phase diagram of CeCo(In$_{1-x}$Cd$_{x}$)$_5$ ($x =0.0075$) \cite{Sunil2010}.}
\end{figure}
Equation (\ref{PhaseBoundary}) also allows us to {\it directly} determine the whole quantum critical phase boundary 
$H_{QCP}(P)$ from our $\rho_{ff}$ measurements done in the mixed state. The antiferromagnetic boundary in the $H-P$ plane at $T=0$, which represents the $H_{QCP}$ line, is shown in Fig. 3(b) (red line). The upper critical field and quantum critical field boundaries merge at $P_c$.  Data points denoted by daggers, open diamonds, and right- and left-hand triangles are data from Ref. [\onlinecite{HQCP}].

We also plot on this figure QCP points previously reported in the literature, as extracted from different experimental techniques. The open circle is QCP determined at zero pressure through both Hall effect \cite{Hall} and thermal expansion \cite{Zaum} measurements. This data point falls onto the presently determined $H_{QCP}(P)$ line.  
The solid diamonds are QCP points extracted by fitting resistivity data with $\rho=\rho_0+a_0(H-H_{QCP})^{-n}$ in which an $n = 1.37$ was chosen in order to obtain $H_{QCP} = H_{c2}(0) = 5.1$ T at ambient pressure \cite{Paglione,RonningTc}. Although these data points do not fall onto the present $H_{QCP}(P)$ line, we find that $n=2$ can also be used to fit very well the same resistivity data, yielding $H_{QCP}(P=0)=4.1$ T, which is the same value as previously reported \cite{Hall}, and the newly obtained $H_{QCP}(P)$  points (solid circles) now follow very well the  red $H_{QCP}(P)$ line, extracted from our present work. The yellow doted line is the  $H_{QCP}$  line that we calculated using Eq. (2) of Ref.  \cite{SachdevVortices} with $H_{c2}(P=P_c)=5.3$ T and our values of $P^*=2.8$ GPa and $P_c=-0.75$ GPa. This theoretical curve overlaps remarkably well with our $H_{QCP}$ line, with a small deviation at high pressure values, which is expected since the theoretical work only gives $H_{QCP}$
close to  $P_c$. Hence, this composite figure provides support for our analysis of the data and their interpretation. It also shows that the flux-flow measurement technique is a powerful tool to probe the subtle physics of the antiferromagnetic phase, which otherwise is undetectable because it is precluded by the superconducting  transition.

We show in Fig. 3(c) the antiferromagnetic and superconducting boundaries in the $H-T$ plane at ambient pressure for CeCoIn$_5$. The open circles are upper critical field data taken from Ref. \cite{Zaum}. Equation (1), which gives the antiferromagnetic boundary, shows that this boundary is suppressed ($T_N$ and $H_N$ are suppressed) with increasing pressure, with the AF boundary being outside the SC dome for $P<P_c$, overlapping with the SC boundary at  $P=P_c=-0.75$ GPa, entering the SC dome for $P>P_c$, and collapsing into the $H=T=0$ point at $P=P^*\approx 2.8$ GPa. The inset shows the $H-T$ phase diagram of CeCo(In$_{1-x}$Cd$_{x}$)$_5$ ($x=0.0075$) taken from Ref. \cite{Sunil2010} that corresponds to a chemical pressure of -1 GPa. This figure is consistent with the just discussed findings based on Eq. (1), showing that, indeed, for a negative pressure smaller than $P_c$, the AF boundary (red stars) is outside the superconducting dome.

To summarize, we observed a sharp increase in quasiparticle scattering inside the vortex core of CeCoIn$_5$ with decreasing $H$ and $T$. We attribute this result to the presence of critical spin fluctuations near $T_N$ inside the vortex core.  This upturn in the vortex core resistivity is significantly suppressed by applied pressure, most likely since the AF order is suppressed with increasing $P$. Based on the scaling behavior of the vortex core resistivity, we identified the AF phase boundary within the SC dome as described by Eq. (1). In essence, these results provides evidence that the microscopic structure of the superconducting phase in CeCoIn$_5$ is highly unusual. The synergy of magnetism and superconductivity gives rise to a composite superconducting state in which conduction and magnetic degrees of freedom are strongly coupled to each other. Our experiment also shows the potential of the flux-flow measurement technique in probing the subtle features of unconventional superconductivity, in particular, how it competes with other phases, and in providing important insight into the nature of the interplay between quantum criticality, magnetism, and superconductivity in other strongly correlated systems such as iron-pnictides and copper oxides, as long as the pinning strength, and thermal and quantum fluctuations are small (see Supplementary material for details). 
     
\begin{acknowledgments}
The authors are grateful to C. Petrovic, N. Curro, A. Bianchi, I. Vekhter, and K. Ueda for useful discussions. This work was supported by the National Science Foundation (grant NSF DMR-1006606 and DMR-0844115),  ICAM Branches Cost Sharing Fund from Institute for Complex Adaptive Matter, and Ohio Board of Regents (grant OBR-RIP-220573) at KSU, and by the U.S. Department of Energy (grant DE-FG02-04ER46105) at UCSD. 
\end{acknowledgments}
$^1$Permanent address: Institute of Physics, Chinese Academy of Sciences, Beijing 100190, China.

\vspace{1cm}
\begin{center}

\newpage

{\large \bf Supporting online material: Strong magnetic fluctuations in the superconducting state of CeCoIn$_5$}\\[0.5cm]

T. ~Hu$^\star$, H. ~Xiao$^{\star,1}$,  T. A. ~Sayles$^\dagger$, M. ~Dzero$^\star$, M. B. ~Maple$^\dagger$, and C. C. ~Almasan$^\star$\\
$^\star${\em Department of Physics, Kent State University, Kent, OH 44242, USA }\\ 
$^\dagger${\em
Department of Physics, University of California at San Diego, La Jolla, CA 92903, USA
}
\date{\today}
\end{center}\vspace{0.5cm}

\section{Experimental methods}
Single crystals of CeCoIn$_5$ were grown using the flux method. High quality crystals  were chosen to perform current-voltage ($I-V$) measurements as a function of temperature ($T$) and applied magnetic field ($H$), with $H \parallel c$  crystallographic axis. The single crystals have a typical size of $2.1\times 1.0 \times 0.16$ mm$^3$, with the $c$ axis along the shortest dimension of the crystals. The single crystals were etched in concentrated HCl for several hours to remove the indium left on the surface during the growth process. They were then rinsed thoroughly in ethanol.
Four leads were attached to the single crystal, with  $I \parallel a$. A small contact resistance (less than 0.3 $\Omega$) is crucial  to keep the heat dissipation in the sample to negligible values during the $I-V$ measurements in applied curents up to 20 mA.  The $I-V$ measurements were performed in a Physical Property Measurement System using an LR700 resistance bridge to apply the excitation current and to measure the longitudinal voltage. The magnetoresistivity was measured both in ambient pressure as well as  under  hydrostatic pressure ($P$) up to 2.35 GPa. The measurements under hydrostatic pressure were perfomed after mounting the single crystal in a pressure cell that uses equal parts of pentane and iso-pentane as pressure transmiting fluid. We applied the desired pressure at room temperature and determined the actual pressure at low temperatures by measuring the superconducting transition temperature of tin for the particular applied pressure.

Typical $I-V$ curves measured in the mixed state of CeCoIn$_5$ at 2.2 K and  0.36 GPa in  different magnetic fields $H \parallel c$ axis are shown in Fig. S1.  
Note that $I_c$  (defined as the extrapolation of the linear $I$-$V$ range to zero voltage) decreases with increasing $H$ due to suppressed pinning energy. The upper critical field ($H_{c2}$) for a particular temperature is defined as the field at which  $I_c$ is zero.  Similar $I-V$ curves were obtained at other temperatures and pressures. 

\setcounter{figure}{0}
\renewcommand{\figurename}{Fig.S}
\begin{figure}
\centering
\includegraphics[trim=0cm 0cm 0cm 0cm, clip=true, width=0.5\textwidth]{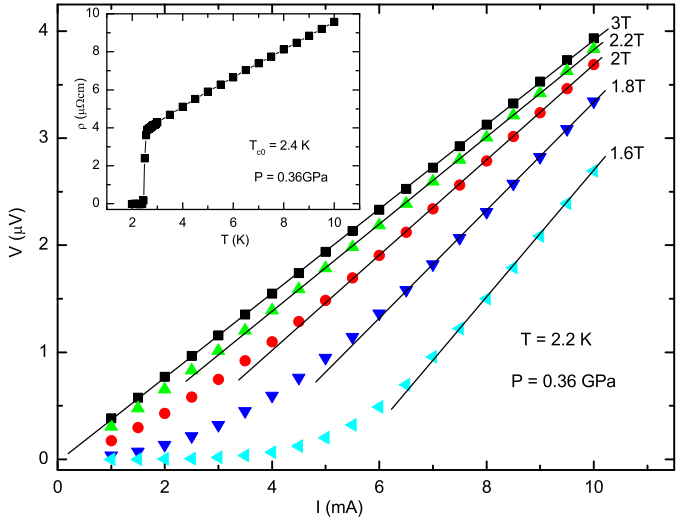}
\caption{ \label{S} (Color online) Voltage $V$ vs current $I$ of CeCoIn$_5$, measured at 2.2 K and 1.6, 1.8, 2.0, 2.2, 3.0 T in a hydrostatic pressure $P$ of 0.36 GPa. Inset: Temperature $T$ dependent resistivity $\rho$ measured in zero field and at 0.36 GPa.}
\end{figure}

On Fig. S2 we present a plot of the $H-T$ phase diagram of CeCoIn$_5$ generated from published zero resistance data  \cite{Paglione} (triangles) and data from this work (circles) as defined above and shown in Figs. 1 of the main text. The shift between the two sets of data is a result of the different definition used for $T_c(H)$ in Ref. \cite{Paglione}. Nevertheless, these two sets of data show that the definition of $H_{c2}$ and $T_c$ used here is correct.

Since the normal-state resistivity of CeCoIn$_5$ displays a significant $T$ dependence (see inset to Fig. S1), any amount of Joule heating would give  a non-linear $I-V$ curve in the normal state. The fact that the measured $I-V$ curves above $H_{c2}$ (see, for example, the 3 T  data of Fig. S1) are linear (Ohmic) over the whole  measured $I$ range shows  that Joule heating  in the sample is negligible. 

As the temperature decreases, the pinning, or the critical current $I_c$, increases. Consequently, 
 higher values of the applied current are required to determine $\rho_{ff}$ from linear $I$-$V$ measurements. However, higher currents give rise to Joule heating and temperature instability  in the cryostat. Therefore, in order to extend the $I$-$V$ measurements to lower $T$ and, at the same time to overcome Joule heating, we performed $I$-$V$ measurements for $T$ down to 1 K at a constant reduced magnetic field $H/H_{c2}=0.94$, i.e. in the regime where the pinning is weak. A plot of the temperature dependence of the normalized flux-flow resistivity, extracted from these $I$-$V$ measurements (inset to top right panel of Fig. 1 in the main text), shows that $\rho_{ff}$ continues to increase with decreasing temperature.
 
 \begin{figure}
\centering
\includegraphics[trim=0cm 0cm 0cm 0cm, clip=true, width=0.5\textwidth]{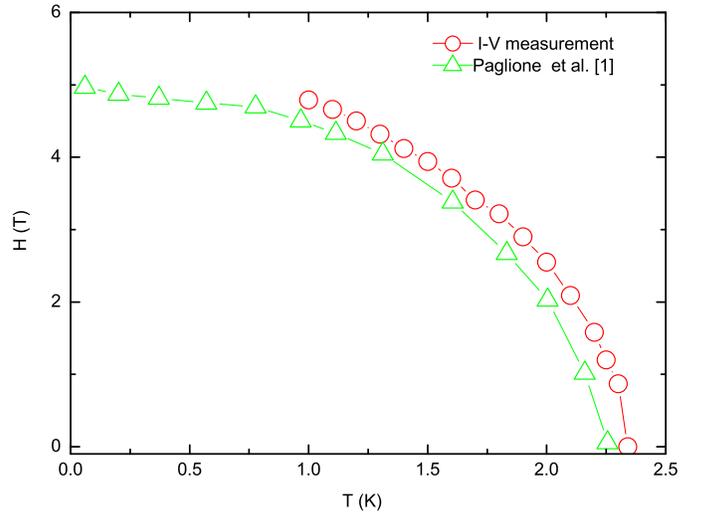}
\caption{ \label{fig:FigS2} (Color online)  $H-T$ phase diagram generated with data from Ref. \cite{Paglione} (triangles) and present work (circles).}
\end{figure}

The linear scaling behavior (that passes through the origin and with a slope that decreases with increasing $P$), revealed by the resistivity  data when plotted as normalized flux-flow resistivity $\Delta\rho_{ff}(H)/\rho_{ff}(H_{c2})$ vs $(\Delta H/H_{c2})^2(\Delta T/T_{c0})^2$  [Fig. 2(b) main text], gives the following scaling law:
 \begin{equation}
\frac{\Delta \rho_{ff}(H,T)}{\rho_{ff}(H_{c2},T)}=A(P)\left(\frac{\Delta T}{T_{c0}}\right)^2\left(\frac{\Delta H}{H_{c2}}\right)^2,
\end{equation} 
where $\Delta \rho_{ff} \equiv \rho_{ff}(H,T)-\rho_{ff}(H_{c2},T)$, $\Delta T \equiv T_{c0}-T$, $\Delta H \equiv H_{c2}-H$, and  $A$ is the pressure dependent slope. 
This equation shows that one can generate contour plots of $\Delta\rho_{ff}(H)/\rho_{ff}(H_{c2})$  in the $H-T$  plane for a fixed $P$. The $\Delta\rho_{ff}(H)/\rho_{ff}(H_{c2})$ = 0 contour (i.e., either $T=T_{c0}$ or $H=H_{c2}$) corresponds to the $H-T$ $SC$ boundary [see Fig. 3(c) of main text]. As $\Delta\rho_{ff}(H)/\rho_{ff}(H_{c2})$ increases (i.e., either $H$ or $T$ decreases), the contour moves deeper under the SC dome. 
Figure S3(a) shows that the parameter $A(P)$ increases with decreasing $P$, with a stronger increase near
ambient pressure and a clear indication of divergence at a slightly negative pressure $P_c$. An extrapolation of these data to high $P$ gives $P^{*} \approx 2.8$ GPa at $A = 0$, the pressure at which the upturn in resistivity, hence AF order, is completely suppressed. 
A plot of $A^{-1/2}(P)$ vs. $P$ [Fig. S3(b)] shows a linear $P$ dependence  such that 
\begin{equation}
A^{-1/2} (P)=A_0^{-1/2}(P-P_c)/(P^{*}-P_c), 
\end{equation}
with $P_c=-0.75$ GPa and $A_0=55$, obtained from a linear fit of these data. In summary, at $P_c$  [$A$ in Eq. (1) diverges], the AF and SC boundaries coincide, while at $P^{*}$ ($A$ becomes zero), the AF boundary collapses into the $H=T=0$ point [see Fig. 3(c)]. 

 \begin{figure}
\centering
\includegraphics[trim=0cm 0cm 0cm 0cm, clip=true, width=0.5\textwidth]{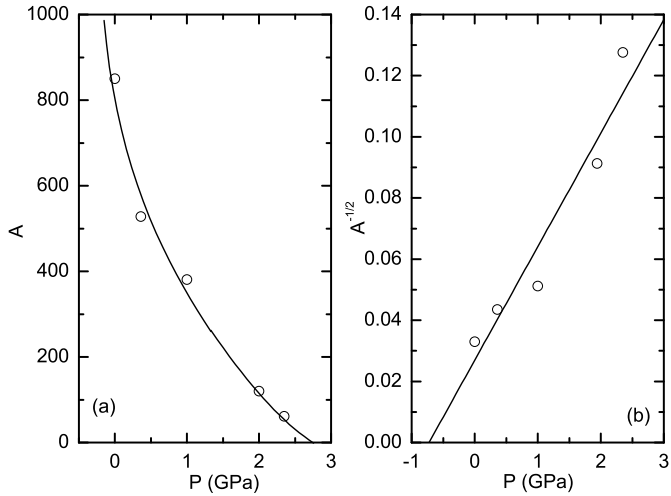}
\caption{ \label{fig:FigS3} (Color online) Plot of slopes of the linear fits $A$ vs $P$ and of $A^{-1/2}$ vs $P$, where parameter $A(P)$ is defined by $\Delta\rho_{ff}(H)/\rho_{ff}(H_{c2})=A(P)f(H,T)$.}
\end{figure}

With the $A(P)$ dependence given by Eq. (2), the present data show that the normalized resistivity follow the expression
 \begin{equation}
\frac{\Delta \rho_{ff}(H,T)}{\rho_{ff}(H_{c2},T)}=A_0\left(\frac{P^{*}-P_c}{P-P_c}\right)^2\left(\frac{\Delta T}{T_{c0}}\right)^2
\left(\frac{\Delta H}{H_{c2}}\right)^2.
\end{equation} 
This shows that $\Delta\rho_{ff}(H)/\rho_{ff}(H_{c2})=A_0$ for the $H=T=0$ AF point corresponding to  $P=P^{*}$. In order to obtain an analytical expression of the AF phase boundary,  we next assume that  $\Delta\rho_{ff}(H)/\rho_{ff}(H_{c2})=A_0$ on any AF boundary, regardless the value of pressure. This assumption is reasonable since  $\Delta\rho_{ff}(H)/\rho_{ff}(H_{c2})$ is a result of critical spin fluctuations, whose value should not depend on $P$.  As a result, The above Eq. (3) gives the phase boundary between the paramagnetic and antiferromagnetic phases inside the superconducting state shown in the main text as Eq. (1).
\section{Numerical estimates}
How well vortices will propagate through the system is mainly determined by the static disorder. Disorder leads to the pinning of the vortices. The strength of the pinning effects can be estimated from the ratio of the critical current density $j_c$ to the pair breaking current density $j_0$ \cite{VortexReview}. Since we are only concerned with an order of magnitude estimate, we will ignore the material's anisotropy and replace the relevant quantities with their averages over the three spacial directions. The value of the pair-breaking current 
is then given by $j_0=c\Phi_0(0)/12\sqrt{3}\pi\lambda_0^2\xi(0)$, where $\lambda_0\sim 400$ nm \cite{PenDepth} is the penetration depth at zero temperature, $\Phi_0=hc/2e$ is the flux quantum, and $\xi(0)\sim 7$ nm is the coherence length \cite{Review1}. In CeCoIn$_5$ at temperatures $T\sim 1$ K and very low magnetic fields  $j_c\sim 4\times 10^3$ A/cm$^2$ \cite{CriticalCurrent}, so that $j_c/j_0\sim 10^{-3}-10^{-2}$. Thus, our estimate shows that pinning forces are very weak in CeCoIn$_5$. 

Static disorder is not the only factor which governs the behavior of the vortex matter. In particular, the effects of thermal and quantum fluctuations may also play an important role \cite{VortexReview}. We recall that the strength of thermal fluctuations is described by the 
Ginzburg number $\textrm{Gi}\sim(1/32\pi^2)(k_B/\xi^3(0)\Delta C)^2$, where $\Delta C\simeq 3\times 10^3$ J/(m$^3\cdot$K) is a volumetric specific heat jump at the superconducting transition \cite{Review1,SpecificHeatJump}. 
Using the experimental data of Refs. \cite{Review1,SpecificHeatJump}, we find $\textrm{Gi}\sim 10^{-7}-10^{-6}$, so that the thermal fluctuations are weak. Finally, let us make an estimate for the quantum resistance $Qu\sim 1/n_c\varepsilon\xi^3(0)$, where $n_c$ is a carrier density and $\varepsilon\sim 0.7$ is an anisotropy parameter. $Qu$ describes the effect of the macroscopic quantum fluctuations on a superconductor due to the vortex motion. We find $Qu\sim 3\times 10^{-3}$, so that the macroscopic quantum fluctuations also appear to be weak. Therefore, from all these estimates we conclude that the proximity of the antiferromagnetic phase has the largest effect on the flux flow resistivity, at least at small to moderate values of pressure and applied magnetic field.

$^1$Permanent address: Institute of Physics, Chinese Academy of Sciences, Beijing 100190, China.

\end{document}